\documentclass[aps,prl,showpacs, twocolumn, superscriptaddress]{revtex4-1}

\usepackage{graphicx}
\usepackage{dcolumn}
\usepackage{bm}
\usepackage{color}
\usepackage{colortbl}
\usepackage{amsmath}
\usepackage{amssymb}
\usepackage{mathrsfs} 

\usepackage{multirow}

\begin{document}

\title{Quantum optical synthesis in 2D time-frequency space}
\author{Rui-Bo Jin}
\affiliation{Hubei Key Laboratory of Optical Information and  Pattern Recognition,  Wuhan Institute of Technology, Wuhan 430205, China\\}
\author{Kurumi Tazawa}
\affiliation{The University of Electro-Communications, 1-5-1 Chofugaoka, Chofu, Tokyo, Japan}
\author{Naoto Asamura}
\affiliation{The University of Electro-Communications, 1-5-1 Chofugaoka, Chofu, Tokyo, Japan}
\author{Masahiro Yabuno}
\affiliation{National Institute of Information and Communications Technology,  588-2 Iwaoka, Kobe 651-2492, Japan}
\author{Shigehito Miki}
\affiliation{National Institute of Information and Communications Technology,  588-2 Iwaoka, Kobe 651-2492, Japan}
\affiliation{Graduate School of Engineering Faculty of Engineering, Kobe University, 1-1 Rokko-dai cho, Nada-ku, Kobe 657-0013, Japan}
\author{Fumihiro China}
\affiliation{National Institute of Information and Communications Technology,  588-2 Iwaoka, Kobe 651-2492, Japan}
\author{Hirotaka Terai}
\affiliation{National Institute of Information and Communications Technology,  588-2 Iwaoka, Kobe 651-2492, Japan}
\author{Kaoru Minoshima}
\affiliation{The University of Electro-Communications, 1-5-1 Chofugaoka, Chofu, Tokyo, Japan}
\author{Ryosuke Shimizu}
\email{r-simizu@uec.ac.jp}
\affiliation{The University of Electro-Communications, 1-5-1 Chofugaoka, Chofu, Tokyo, Japan}

\date{\today }

\begin{abstract}
\textbf{Conventional optical synthesis, the manipulation of the phase and amplitude of spectral components to produce an optical pulse in different temporal modes, is revolutionizing ultrafast optical science and metrology.
These technologies rely on the Fourier transform of light fields between time and frequency domains in one-dimensional space. However, within this treatment it is impossible to incorporate the quantum correlation  among photons.
Here we expand the Fourier synthesis into high dimensional space  to deal with the quantum correlation, and carry out an experimental demonstration by manipulating the two-photon probability distribution of a biphoton in two-dimensional time and frequency space.
As a potential application, we show  manipulation of a heralded single-photon wave packet, which is never explained by the conventional one-dimensional Fourier optics.
Our approach opens up a new pathway to tailor the temporal characteristics of a biphoton wave packet with high dimensional quantum-mechanical treatment.
We anticipate such high dimensional treatment of light in time and frequency domains could bridge the research fields between quantum optics and ultrafast optical measurements.
}

\end{abstract}



\maketitle

\section{Introduction}

The invention of mode-locked lasers opened the door for ultrafast optical science and technology in the femtosecond region. This field has been continuously developed by manipulating the temporal waveform of optical pulses through the one-dimensional (1D) Fourier transform relationship of electric field distributions between the time and frequency domains \cite{Weiner2009book, Cundiff2001}.
Today, these technologies are known as ``optical synthesis'' (OS),  which can precisely control and generate  arbitrary temporal shapes of optical pulses by manipulating the phase and  amplitude of  spectral components \cite{Jiang2005, Chan2011, Spencer2018}.
Such OS technologies are expected to be applied to optical measurement, sensing, and spectroscopy \cite{Cundiff2001, Giorgetta2010, Kato2017SRminoshima, Asahara2019APEminoshima}.

From the viewpoint of the particle nature of light, the 1D Fourier transform treatment is valid only for an ensemble of photons without quantum correlation.
On the other hand, recent developments in quantum optical technologies allow us to observe the time-frequency behavior of quantum-mechanically correlated photon pairs, e.g., biphotons \cite{Peer2005}.
It has been revealed that biphoton distributions in the time and frequency domains are connected to Fourier optical phenomena in two-dimensional (2D) time-frequency space \cite{Jin2018PRAppl, MacLean2018, Ansari2018, Ansari2018Optica, Davis2018arXiv, Graffitti2019arXiv}.
Such a quantum optical aspect of light has the potential to expand the conventional optical synthesis to its quantum counterpart, i.e.,
optical synthesis in high-dimensional time-frequency space. This treatment allows us to incorporate the quantum correlation into the OS. Hereafter, we refer to  OS in high-dimensional space  in a quantum manner as ``quantum optical synthesis'' (QOS).

In this work, we present a proof-of-concept demonstration of QOS by manipulating the amplitude and phase of the spectral distribution of biphotons in 2D  frequency space, and by directly observing the temporal distribution in 2D time space.
Furthermore, as a potential application of QOS, we show the shaping of heralded single photons via manipulation in 2D time-frequency space.
%
%
%
%
\begin{figure*}[tbhp]
\centering
\includegraphics[width= 0.55\textwidth]{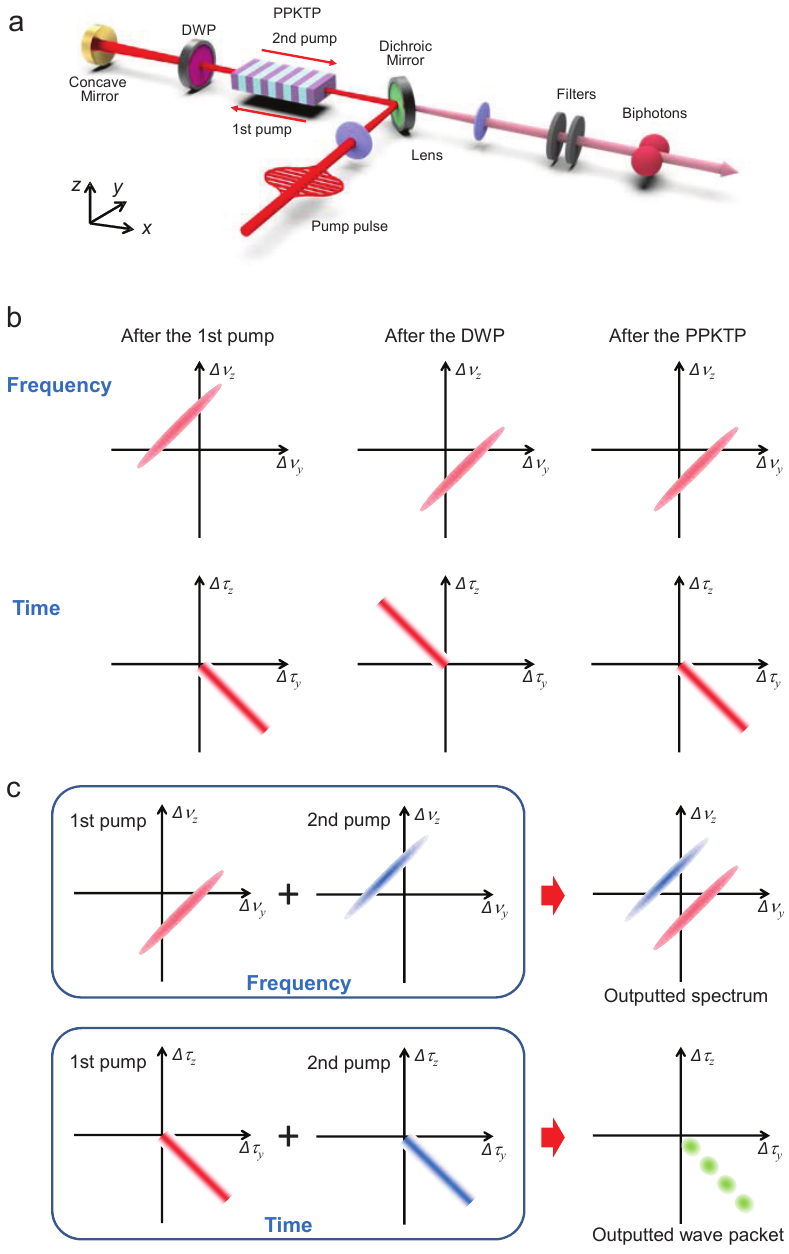}
\caption{ \textbf{Experimental scheme for manipulating a biphoton wave packet in 2D time-frequency space. }
(a) Schematic drawing of a bidirectional pumping. A pair of photons with orthogonal polarization is generated in  either the first (1st) or second (2nd) pumping for the PPKTP via spontaneous parametric down-conversion. (b) Expected two-photon spectral distributions (upper) and corresponding temporal distribution with respect to a pump pulse position (lower).  The upper and lower figures each represent a two-photon distribution just after the 1st  pumping (left), after passing through the DWP twice and before the 2nd pumping (center), and after the 2nd pumping (right). (c) Superposition of the biphoton wave packets produced by the 1st (red) and 2nd (blue) pumpings. Two separated modes can be observed in the frequency domain, but the modes completely overlap in the time domain. This temporal overlapping results in a modulation of the biphoton  wave packets in 2D time space.
}
\label{concept}
\end{figure*}
%

%
%
%
\begin{figure*}[!t]
\centering
\includegraphics[width= 0.65\textwidth]{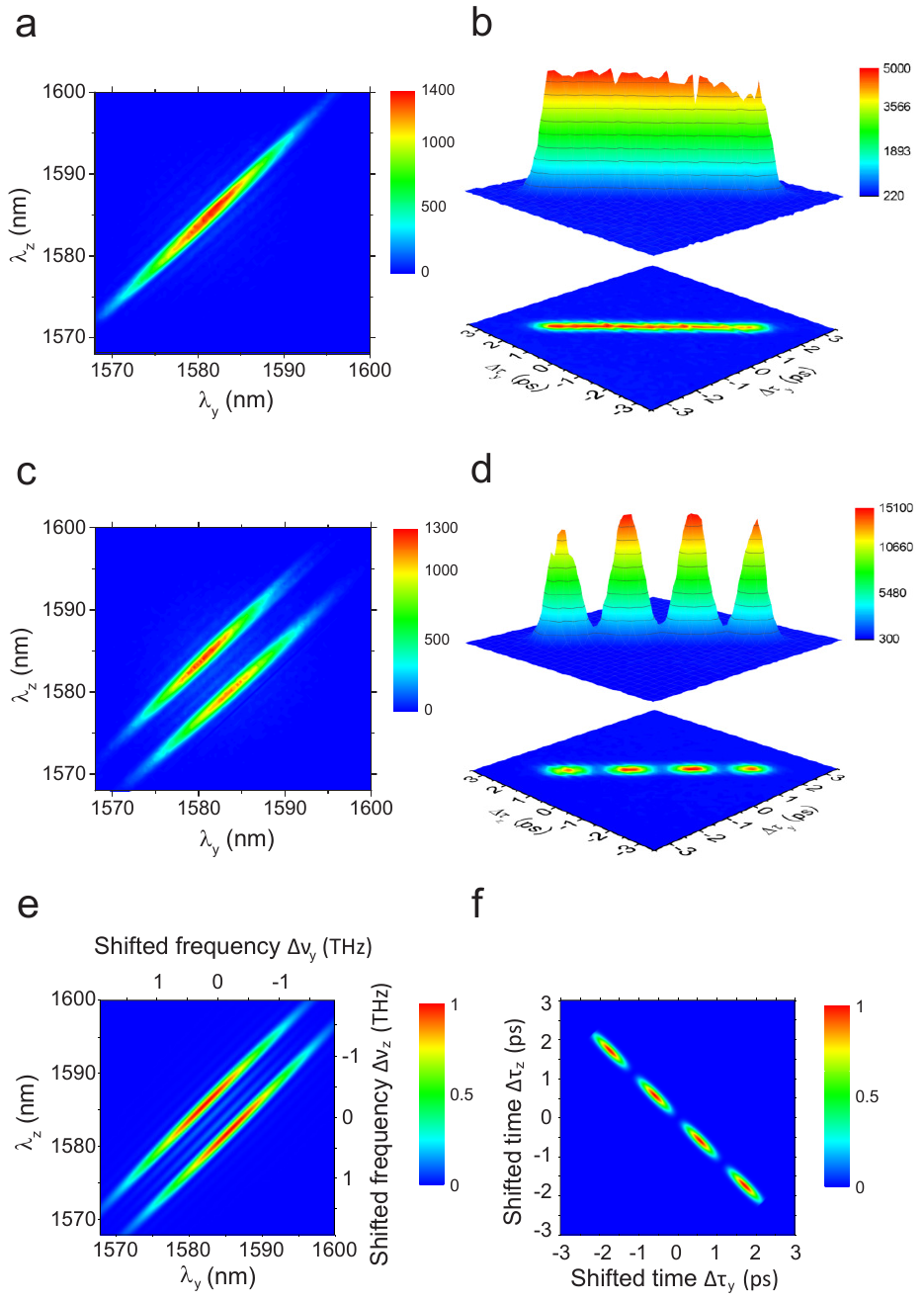}
\caption{  \textbf{Experimental demonstration of quantum optical synthesis by manipulating the number of spectral modes.}
(Upper) Experimentally measured JSI (a) and JTI (b) with a single-pumped SPDC at the crystal temperature of 65 $^\circ$C.
(Middle) Experimentally measured JSI with the bidirectional pumping scheme (c) and the corresponding JTI (d).
(Lower) Theoretically simulated JSI assuming the PPKTP length of 30 mm and the pump pulse spectral width of 7.4 nm (e). We set the mode separation of 0.58 THz so as to reproduce the experimental data. Figure (f) is the JTI obtained from the Fourier transformation of the joint spectral amplitude for the left figure.
}
\label{amplitude}
\end{figure*}
%

\section{Experiment and Results}
The scheme of our QOS experiment is shown in Fig.\,\ref{concept}.
Laser pulses with a center wavelength of 792 nm and a bandwidth of 7.4 nm were used to pump a 30-mm-long PPKTP crystal in a bidirectional pumping configuration \cite{Branning1999, Jin2018OE}, as shown in Fig.\,\ref{concept}a.
The PPKTP crystal with type-II phase matching satisfies the group-velocity-matching (GVM) condition at the telecom wavelength  \cite{Grice2001, Eckstein2011,  Cui2019PRAppl}.
The polarization of the constituent photons is aligned along either the crystallographic y or z axis.
Thanks to the GVM condition with femtosecond pulse pumping \cite{Shimizu2009}, the down-converted biphotons have an elliptical distribution along the diagonal direction in the frequency domain  and along the  anti-diagonal direction in the time domain, as shown on the left of Fig.\,\ref{concept}b.
After passing twice through the dual waveplate (DWP) (quarter-wave plate for the biphotons and half-wave plate for the pump), the y(z)-polarized photon is interchanged with a z(y)-polarized photon, but  the pump pulse is unchanged.
As a result, the two-photon spectral (temporal) distribution is inverted with respect to $\Delta \nu_z = \Delta \nu_y$ $(\Delta \tau_z = \Delta \tau_y)$, as shown in the center of  Fig.\,\ref{concept}b. Here $\Delta \nu$ is defined as the shifted frequency from the center frequency of 189.39 THz (1584 nm), and $\Delta \tau$ is the shifted time from 0 defined by the temporal position of the pump pulse.
Then, only the temporal distribution is shifted along $\Delta \tau_z = -\Delta \tau_y$ after passing through the PPKTP crystal again, as shown on the right of Fig.\,\ref{concept}b.
By combining the biphoton wave packet produced by the 1st and 2nd pumpings, a separated two-mode frequency distribution $s_1$ and $s_2$ was prepared, as shown in Fig.\,\ref{concept}c.
Since the separation and relative phase between the two frequency modes would directly affect the two-photon temporal distribution, we could expect  to synthesize two-photon temporal distributions  by means of two-photon spectral manipulation.
%

%
%
\begin{figure*}[!t]
\centering
\includegraphics[width= 0.8\textwidth]{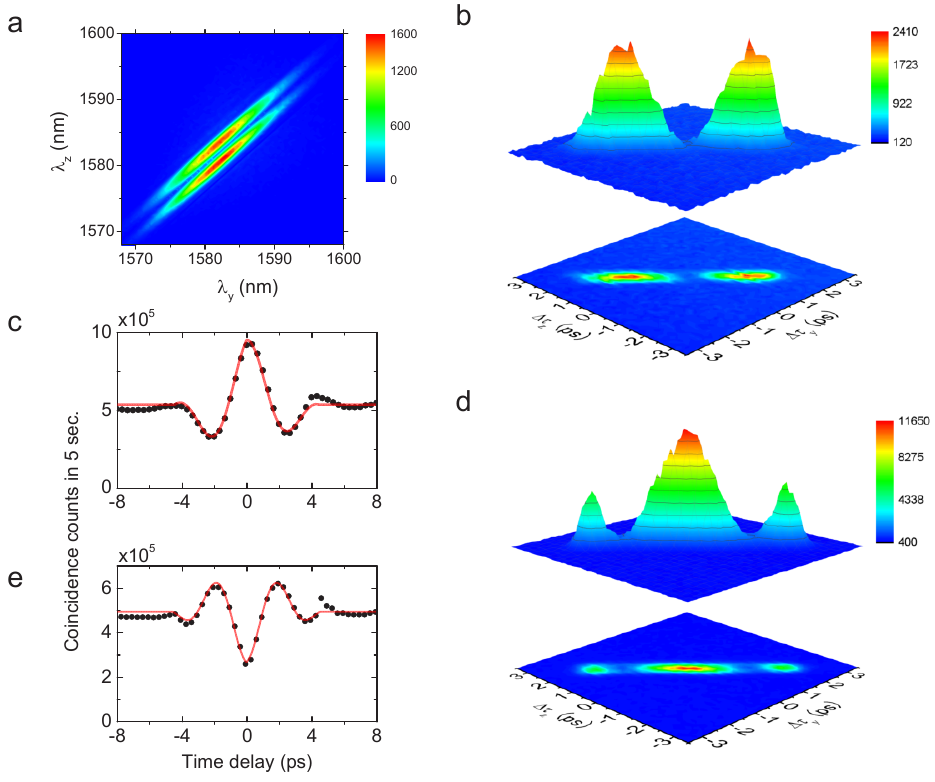}
\caption{  \textbf{Experimental demonstration of quantum optical synthesis by manipulating the relative phase. }
(a) Experimentally measured JSI at the crystal temperature of 45 $^\circ$C. (b) Corresponding JTI without the glass plate in order to control the relative phase. (c) Hong-Ou-Mandel interference patterns in the condition of (b). (d) JTI with the phase manipulation by inserting and tilting a silica glass plate with a thickness of 1.5 mm. (e) Hong-Ou-Mandel interference patterns in the condition of (d).
}
\label{phase}
\end{figure*}
%

%
%
%
Based on the scheme described above, we performed proof-of-principle experiments for QOS from two aspects: (1) the multiple mode effect in 2D frequency space and (2) the phase manipulation effect.
These effects were confirmed by directly measuring two-photon spectral and temporal distributions in 2D space, where the 2D spectral and temporal distributions are known as the joint spectral intensity (JSI) and the joint temporal intensity (JTI) of biphotons, respectively.
The JSI was measured by a fiber spectrometer \cite{Gerrits2015, Jin2016QST} and the JTI was measured by using a time-resolved up-conversion system \cite{Kuzucu2008PRL, MacLean2018, Jin2018PRAppl}. See the  Supplementary file for the detailed experimental setup.
To identify the multiple mode effect, we begin by presenting the JSI and JTI of the biphotons with a single pumping, realized by placing a filter that blocks biphotons but passes the pump pulse instead of the DWP,  as shown in Fig.\,\ref{amplitude}a and b.
Here we set the crystal temperature to  65 $^\circ$C.
The frequency (time) of the biphotons in JSI (JTI) was positively (negatively) correlated, and no modulations were observed in the JTI.
In contrast, changing the experimental configuration to a bidirectional pumping scheme,  we can clearly see two mode spectral distributions with a positive frequency correlation (Fig.\,\ref{amplitude}c), where the resultant peak separation between the two modes is 4.89  nm (0.58 THz) .
Here, the peak separation is defined as the distance along the line of $\Delta \nu _z =-\Delta \nu _y$.
On the other hand, the observed JTI shows a negative correlation with almost the same full length as that in Fig.\,\ref{amplitude}b, but it has  a distinct four-mode structure along the anti-diagonal direction with a peak separation of 1.60 ps between modes.
The experimental peak separation of 0.58 THz in the JSI and that of 1.60 ps in the JTI satisfy an almost inverse relationship: 0.58$\times$1.60 = 0.93; this suggests that the time and frequency distributions could have a conjugate relationship and partially validates our concept of QOS.
In order to confirm the Fourier transform relationship between the data in  Fig.\,\ref{amplitude} c and those in d, we theoretically constructed the two-photon spectral distribution (Fig.\,\ref{amplitude}e) by adopting the experimental parameters of the pump bandwidth and crystal length.  We then obtained the simulated JTI (Fig.\,\ref{amplitude}f) by performing a two-dimensional Fourier transformation on the amplitude of the JSI in Fig.\,\ref{amplitude}e.
This theoretical simulation can well reproduce the experimental data, meaning it can successfully demonstrate our concept of QOS. The slightly ``fatter'' distribution in  Fig.\,\ref{amplitude}c is caused mainly by the limited temporal resolution of our systems.
%

%
%
\begin{figure*}[!t]
\centering
\includegraphics[width= 0.9\textwidth]{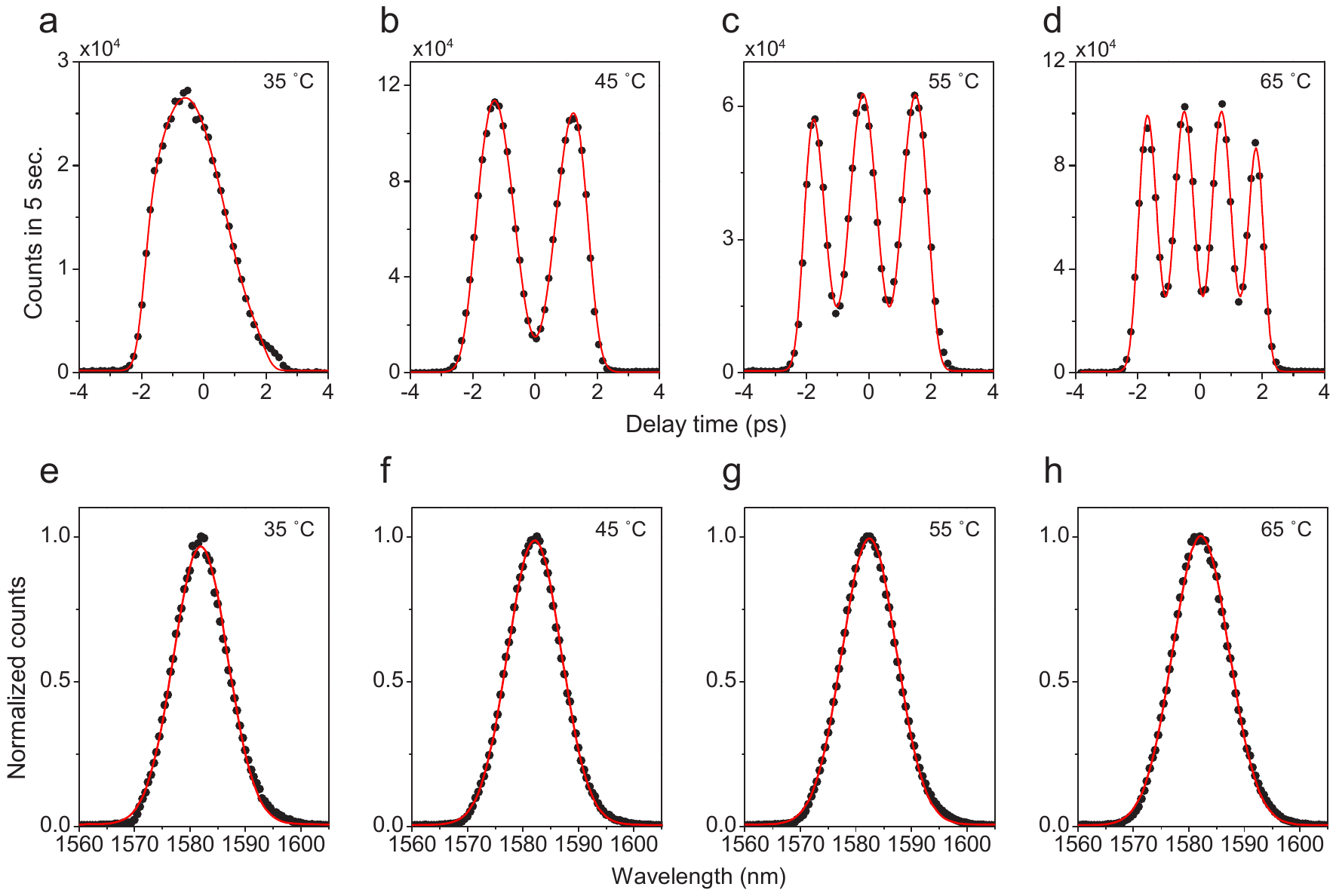}
\caption{  \textbf{Manipulation of heralded single-photon wave packets via Fourier synthesis in 2D time-frequency space.}
The temporal shape of the heralded single-photon wave packets at the  crystal temperature of 35 (a), 45 (b), 55 (c), and 65 $^\circ$C (d) has the peak  numbers of  1, 2, 3, and 4, respectively.
In contrast, the  spectral shape of the heralded single-photon wave packets as shown in (e-h) is basically not changed at different temperatures.
}
\label{application}
\end{figure*}
%

%
To strengthen the  QOS, we verified how the phase manipulation of the biphoton in the frequency domain affects  the two-photon temporal distribution.
First, we decreased the crystal temperature from 65 to 45 $^\circ$C and obtained the mode peak separation of 2.19 nm (0.26 THz) in Fig.\,\ref{phase}a.  The resultant JTI in Fig.\,\ref{phase}b shows a distinct two-mode distribution with a peak separation of 3.41 ps.
The disappearance of the coincidence counts around the 0-delay position, i.e.,
$\Delta \tau _y  = \Delta \tau _z  = 0$,
suggests the relative phase $\phi$ between the two spectral modes in Fig.\,\ref{phase}a to be $\sim\pi$.
In order to ensure the relative phase value, we also carried out the Hong-Ou-Mandel (HOM) interference experiment \cite{Hong1987}  and obtained the interference pattern in Fig.\,\ref{phase}c.
From the fitting, we extracted the relative phase value as $\phi= 0.86\pi$, which approximately agrees with the phase estimated in the JTI.
Next, we controlled the relative phase by inserting and tilting a silica glass plate with a thickness of 1.5 mm between the DWP and the concave mirror in the setup and then obtained the JTI in Fig.\,\ref{phase}d.
The maximum coincidence counts of this JSI are around $\Delta \tau _y  = \Delta \tau _z  = 0$, implying the relative phase $\phi= 0$.
Indeed, we obtained the HOM pattern reflecting the relative phase $\phi= 0$ (Fig.\,\ref{phase}e).
On the other hand, the modulation intervals are almost the same between Fig.\,\ref{phase}b and Fig.\,\ref{phase}d,
meaning that the relative phase $\phi$ only affects the phase of the modulation in the JTI.
The 2D Fourier transformation  well explains all the phenomena between the JSI and JTI.

Next, we discuss the advantages of our manipulation methods.
In our bidirectional pumping scheme, the  whole
two-photon spectral distribution $S$ is expressed by  the superposition of the two-photon spectral modes $s_1$ and $s_2$;
 $S(\omega _1 ,\omega _2 ) = s_1 (\omega _1 ,\omega _2 ) + e^{i\phi } s_2 (\omega _1 ,\omega _2 )$,
where  $s_2 (\omega _1 ,\omega _2 )= s_1 (\omega _2 ,\omega _1 )$  in our case.
The spectrum of the constituent photon with $\omega_{1(2)}$ is given by
$f_{1(2)} (\omega _{1(2)} ) = \int  S (\omega _1 ,\omega _2 )d\omega _{2(1)}$.
In order to control the two-photons spectral distribution  $S(\omega _1 ,\omega _2 )$, we need to independently control the spectral modes  $s_1$ and  $s_2$.
So far, the spatial light modulator (SLM) has been frequently used and enables us to arbitrarily shape an ultrafast optical pulse.
In previous reports, this technique is also applied to quantum optical experiments to control a biphoton temporal waveform \cite{Peer2005, MacLean2018, Lu2018OL}.
In those experiments, the SLM manipulates the spectral amplitude and phase distribution of the constituent photons $f_1$ or $f_2$.
However, the SLM cannot be utilized in our experimental situation because a certain frequency component in the one-photon spectrum  $f_1$ may contain the biphoton spectral modes of both  $s_1$ and  $s_2$.
In contrast, our bidirectional pumping scheme is based on  manipulation along a difference-frequency axis, allowing the independent control of $s_1$ and  $s_2$ even in the spectral condition in Figs.\,\ref{amplitude} and \,\ref{phase}.

In addition to the advantage of manipulation in the difference-frequency axis as discussed above, our scheme can offer a new approach to shape a heralded single-photon wave packet via  manipulation of a high-dimensional time-frequency space.
To clarify this point, we show the manipulation of a single-photon temporal waveform in a heralding scheme. In this experiment, z-polarized photons were detected by the superconducting nanowire single photon detector (SNSPD) without   time-resolved measurement, while the y-polarized photons were sent to the time-resolved up-conversion system. With this setup, we measured the temporal shapes of the heralded single photons at the crystal temperatures of 35, 45, 55, and 65$^\circ$C.
Figure\,\ref{application}(a-d) shows the  counts of the y-polarized photons triggered by the heralding signal from the SNSPD.
We can clearly see  the distinct modulation and observe the increase in the numbers of peaks with the increase in the crystal temperature, manifesting the advantage of our scheme in comparison with the earlier works on biphoton temporal manipulations with the SLM.
However, the corresponding spectra in Figure\,\ref{application}(e-h) are not changed at different temperatures.
This implies that there is no direct connection in the Fourier transform operation in 1D space and the 2D treatment is essentially required for understanding the time-frequency behavior even for the heralded single-photon wave packet.

\section{Discussion}
It is worthwhile to discuss the plausible future direction of the QOS technique.
The main motivation of our work is to bridge the research fields between quantum optics and ultrafast optical measurements; especially for spectroscopy with ultrafast lasers, which has been utilized for the investigation of the dynamical process in physical, chemical and biological materials \cite{Schmitt2007,  Maiuri2020}.
Indeed, spectroscopy with quantum light is attracting much attention as emerging quantum technology \cite{Mukamel2020JPB, Schlawin2013}.
For instance, a recent theoretical work predicts that heralded single photons produced by SPDC could emulate sunlight conditions and that such photons could be used to investigate the dynamical process in complex molecules \cite{Fujihashi2019arXiv}.
Although the arbitrary shaping of temporal modes is crucial for investigating the light-matter interactions in such complex systems, it is hard to control the temporal characteristics of thermal light such as  sunlight because of its incoherent properties.
Quantum optical synthesis in high-dimensional time-frequency space could provide thermal light with high controllability and may contribute to deeper insights into complex molecular systems such as photosynthetic materials.
Another important application of our QOS technique might be time-frequency quantum key distribution \cite{Xu2019arXiv}, because the temporal distributions with negative correlation in Fig.\,\ref{amplitude} and \ref{phase} can be interpreted as  time bins of entangled photon pairs.
Such a high-dimensional quantum key distribution scheme could provide higher key rate  and higher tolerance to noise.

\section*{Acknowledgments}
We thank Prof. Akihito Ishizaki and Prof. Zheshen Zhang for the helpful discussions. This work was supported by the MEXT Quantum Leap Flagship Program (MEXT Q-LEAP) (Grant Number JPMXS0118069242), by JSPS KAKENHI (Grant Numbers JP18H05245 and JP17H01281), JST CREST (Grant Number JPMJCR1671) and by the National Natural Science Foundations of China (Grant Numbers  91836102 and 11704290).

\onecolumngrid
\clearpage

\renewcommand\thefigure{S\arabic{figure}}
\setcounter{figure}{0}

\setcounter{equation}{0}
\renewcommand\theequation{S\arabic{equation}}

\section*{Supplementary Material}

\subsection*{S1: Experimental setup}
%
\begin{figure}[!h]
\centering
\includegraphics[width= 0.65\textwidth]{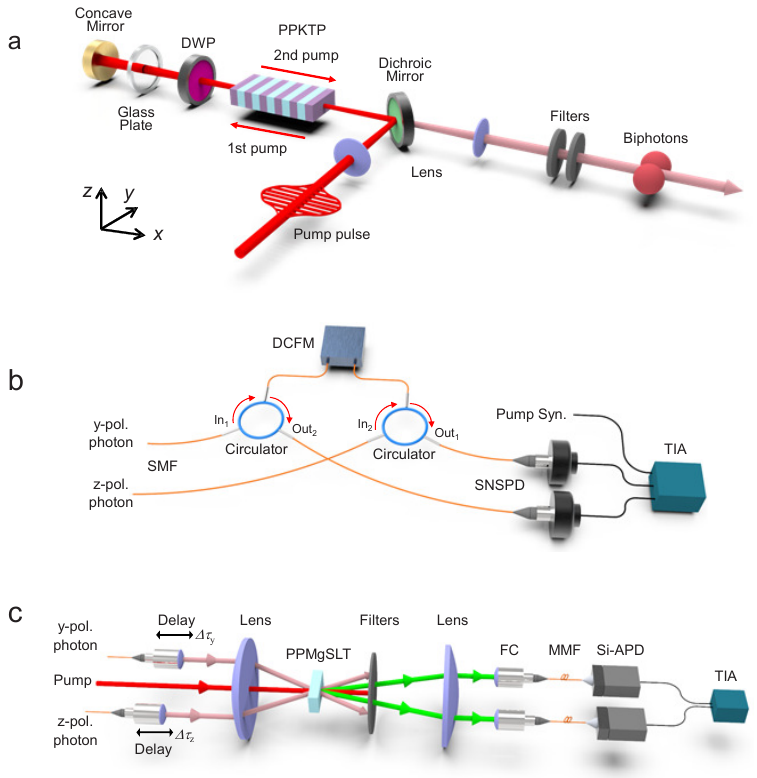}
\caption{  \textbf{Experimental setup for quantum optical synthesis.}
a is the experimental setup for  generating biphotons.
b is the experimental setup for measuring the JSI.
c is the experimental setup for measuring the JTI. DWP, dual wavelength plate; SMF, single-mode fiber; DCFM, dispersion compensation fiber module; SNSPD, superconducting nanowire single photon detector; TIA, time-interval analyzer; Pump Syn., synchronization electrical signal from the pump laser; FC, fiber coupler;  MMF, multi-mode fiber; Si-APD, silicon-avalanched photo diode.
}
\label{FigS-setup}
\end{figure}
%

Figure\,\ref{FigS-setup}a  shows the details of the setup in our experiment.
The JSI was measured by a fiber spectrometer  \cite{Gerrits2015, Jin2016QST}, as shown in Fig.\,\ref{FigS-setup}b. The spectrometer consists of a dispersion compensation fiber module (DCFM), two fiber circulators, two superconducting nanowire single photon detectors (SNSPDs) and a time-interval analyzer (TIA) with coincidence measurement.
The DCFM (from Furukawa Co.) is designed to compensate for the dispersion of a 15-km-long commercial fiber. The DCFM  has a dispersion of -255 ps/nm at 1550 nm  and an insertion loss of 1.3 dB.
Our SNSPDs have a system detection efficiency of around 70\% with a dark count rate less than 100 cps \cite{Miki2013}.
The JTI was measured by using a time-resolved up-conversion system \cite{Kuzucu2008PRL, MacLean2018, Jin2018PRAppl}, as shown in Fig.\,\ref{FigS-setup}c.
For JSI measurement, the pump power for SPDC is ~100 mW.
For JTI measurement, the pump power for SPDC is ~200 mW and the gating power for the JTI is ~400 mW.
The system jitter time is estimated as 123 ps, including jitter time from SNSPD, TIA and the laser trigger signal as 68 ps, 100 ps and 20 ps, respectively. By considering the dispersion of the DCFM as 255 ps/nm at 1550nm, the resolution of our fiber spectrometer is approximately 0.5 nm.
The resolution of the JTI measurement system is estimated to be 340 fs for single-photon measurement and 480 fs for two-photon measurement.
The glass plate and the DWP were removed in order to measure the JSI and JTI of the single-mode  distribution in Fig. 2a and 2b.
A DWP was inserted to measure the two-mode distribution in Fig. 2c and 2d.
A glass plate was inserted  to manipulate the phase in Fig. 3.

\subsection*{S2: Temperature dependence in HOM interference}
\begin{figure*}[!h]
\centering
\includegraphics[width= 0.85\textwidth]{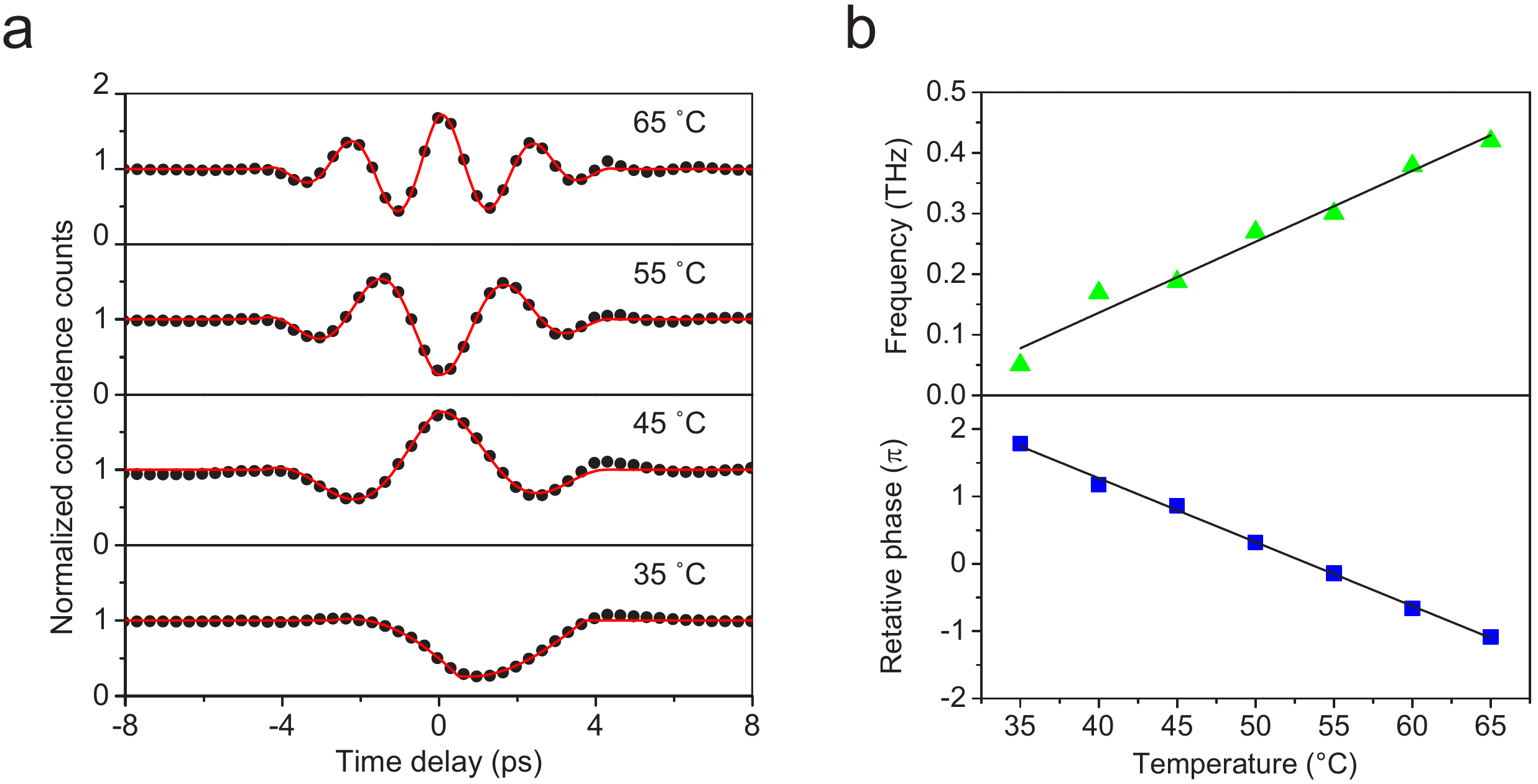}
\caption{  \textbf{Temperature dependence of the beating frequency and the relative phase without the glass plate between two spectral modes.}
(a) Hong-Ou-Mandel interference patterns in different crystal temperatures. We can extract the oscillation periods and relative phase information from the fittings. (b) Plot of the oscillation frequencies (upper) and the relative phase values (lower) as a function of the temperature in PPKTP.
}
\label{FigS-pattern}
\end{figure*}

The relative phase $\phi$  discussed in the main text is governed by the relationship between the pump pulse phase $\phi_p$ and the sum of the phase of the two downconverted biphotons $\phi_y +\phi_z$  before the second pumping, where $\phi_y (\phi_z)$  is the phase of the $y(z)-$polarized photon.
Thus, $\phi$ is controllable through the material dispersion, which is the difference of the refractive indices  between the pump pulse and the biphotons.
In Fig.\,\ref{phase}, we controlled the relative phase $\phi$ by using the dispersion of a silica glass plate with a thickness of 1.5 mm.
In contrast, due to the phase-matching condition, i.e., $\Delta k= k_p-k_y-k_z$, the phase of the pump pulse $\phi_p$ would have some relation to $\phi_y +\phi_z$.
Here $k_p$ is the wave vector of the pump pulse, and $k_y (k_z)$ is that of the y(z)-polarized constituent photon of the biphoton wave packet.
Therefore, changing the PPKTP temperature  changes not only the spectral separation but also  the relative phase $\phi$.
To check the spectral and phase controllability, we measured the HOM interference patterns with the variation of the PPKTP temperature (Fig.\,\ref{FigS-pattern}a) and found that the oscillation frequency increases with the increase in temperature.
By fitting these patterns, we can extract  information on the spectral separation and the relative phase values.
Figure\,\ref{FigS-pattern}b shows the oscillation frequency (upper) and the relative phase (lower) as a function of the PPKTP temperature. Fitting with the linear functions, we found that spectral separation occurs in  $1.2 \times 10^{-2}$ ($THz$/$^{\circ}$C) and that the relative phase value is shifted by  $-9.5 \times 10^{-2}$ ($\pi$/$^{\circ}$C).
The relative phase values are also affected by all the dispersive media between the PPKTP and the concave mirror, the glass plate, the wave plate, and even the air. The dispersion of air is much smaller than that of the glass plate, but the propagation length  is much longer in air than in the other materials. Therefore, the relative phase was also changed by an optical realignment due to the change in length between the PPKTP and the concave mirror.

%
%
%

\newpage

\subsection*{S3: Theoretical simulation of the JSI and JTI }
\begin{figure}[tbhp]
\centering
\includegraphics[width= 0.65\textwidth]{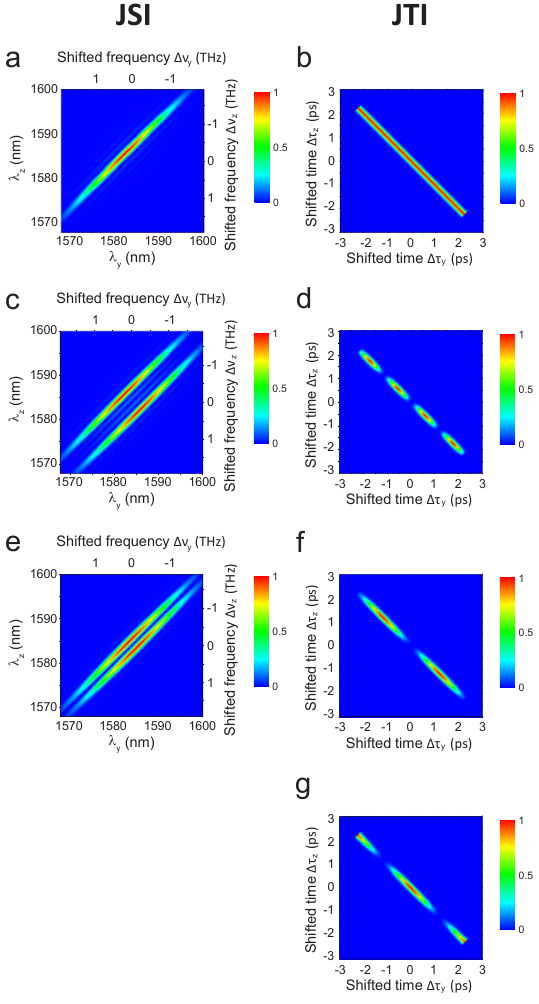}
\caption{  \textbf{Theoretical simulation for the experimental JSI and JTI in the main text.}
a (b)is for Fig. 2a (b).
c (d) is Fig. 2c (d).
e (f) is for Fig. 3a (b).
g is for Fig. 3d.
}
\label{FigS-jsti}
\end{figure}

In this section, we show the procedures to theoretically simulate the experimental joint spectral intensity (JSI) and joint temporal intensity (JTI) in the main text.
The joint spectral amplitude (JSA) $f(\nu_1, \nu_2)$  is the product of the pump envelope function  $\alpha (\nu_1 + \nu_2)$ and the phase matching function   $\phi (\nu_1, \nu_2)$.
\begin{equation}\label{eqS1}
f(\nu_1, \nu_2)=\alpha (\nu_1 + \nu_2)\times \phi (\nu_1, \nu_2)
\end{equation}
where $\nu$ is the linear frequency; the subscripts $1$ and $2$ indicate the down-converted biphotons.
$\alpha (\nu_1 + \nu_2)$  with a Gaussian distribution can be written as
\begin{equation}\label{eqS2}
\alpha(\nu_1+ \nu_2)=\exp[-\frac{1}{2}\left(\frac{\nu_1+\nu_2-\nu_p}{\sigma_p}\right)^2],
\end{equation}
where $\nu_p$ and $\sigma_p$ are the frequency and bandwidth of the pump photons, respectively.
By assuming a flat phase distribution, $\phi (\nu_1, \nu_2)$ can be written as
\begin{equation}\label{eqS3}
\phi(\nu_s,\nu_i)={\rm sinc}\left(\frac{\Delta kL}{2}\right),
\end{equation}
where $L$ is the length of the crystal; $\Delta k=k_p-k_1-k_2+\frac{2\pi}{\Lambda}$ is the difference between the wave vectors; $\Lambda$ is the poling period.
In Fig.\,\ref{FigS-jsti}a,  the JSI of $|f (\nu_1, \nu_2)|^2$  is calculated by considering the pump pulses with a center wavelength of 792 nm and FWHM of 7.4 nm; the PPKTP crystal length was 30 mm.
Figures\,\ref{FigS-jsti}c and \ref{FigS-jsti}e were plotted with the model of $|f (\nu_1, \nu_2)- f (\nu_2, \nu_1)|^2$.
Although the spectral mode separation was caused by the change in crystal temperature, we  effectively control it by changing the poling period. The effective poling period  was obtained by fitting the experimental JSIs in Fig. 2c and Fig. 3a.

Next, we show how to simulate the JTI. For simplicity in the Fourier transformation, the model of $f(\nu_1, \nu_2)$  can be further simplified using a simpler equation:
\begin{equation}\label{eqS4}
JSA(\nu_1, \nu_2)=\exp[-a(\nu_1+\nu_2)^2] \times {\rm sinc} [b(\nu_1-\nu_2)].
\end{equation}
Here, $\nu_1$  and $\nu_2$ are the shifted frequencies, and correspond to $\delta \nu_y$  and $\delta \nu_z$, respectively.
By fitting the data in Fig. \ref{FigS-jsti}a, we obtain the values  a=0.11284 and b=13.888 for this new model. These two parameters are used for all the equations hereafter.
The JTA of $JSA(\nu_1, \nu_2)$ can be calculated using 2D Fourier transformation (FT):
\begin{equation}\label{eqFT}
JTA(t_1, t_2)=FT\{JSA(\nu_1, \nu_2)\} \equiv  \int {\int_{ - \infty }^{ + \infty } { JSA(\nu_1, \nu_2) \exp ( - i2\pi t_1 \nu _1  - i2\pi t_2 \nu _2 )d\nu _1 d\nu _2 } }.
\end{equation}
After a long calculation,
\begin{equation}\label{eqS5}
JTA(t_1, t_2)=\frac{\pi}{2b}\sqrt{\frac{\pi}{a}} \exp[-\frac{\pi^2(t_1+t_2)^2}{4a}] {\rm UnitBox}[\frac{\pi(t_1-t_2)}{2b}],
\end{equation}
where the function of UnitBox is defined as
\begin{equation}\label{eqS6}
{\mathop{\rm UnitBox}\nolimits} (x){\rm{ = }}\left\{ {\begin{array}{*{20}c}
   {{\rm{1,}}\begin{array}{*{20}c}
   {\begin{array}{*{20}c}
   {} & { - \frac{1}{2} \le x \le \frac{1}{2}}  \\
\end{array}}  \\
\end{array}}  \\
   {{\rm{0,}}\begin{array}{*{20}c}
   {\begin{array}{*{20}c}
   {\begin{array}{*{20}c}
   {}  \\
\end{array}} & {{\rm{otherwise}}}  \\
\end{array}}  \\
\end{array}}  \\
\end{array}} \right.
\end{equation}
Figure\,\ref{FigS-jsti}b is plotted using Eq.\,\ref{eqS5}.
For two spectral modes with a relative phase of $\pi$ (differential frequency), the JSA can be expressed by the following equation:
\begin{equation}\label{eqS7}
JSA_-(\nu_1, \nu_2)=\exp[-a(\nu_1+\nu_2)^2] \{{\rm sinc} [b(\nu_1-\nu_2+\Delta)]- {\rm sinc} [b(\nu_1-\nu_2-\Delta)]\},
\end{equation}
where $\Delta$ represents the mode  separation, which can be obtained by fitting the experimental JSI data.
The corresponding JTA can be calculated using 2D FT.
\begin{equation}\label{eqS8}
JTA_-(t_1, t_2)=\frac{i \pi}{b}\sqrt{\frac{\pi}{a}} \exp[-\frac{\pi^2(t_1+t_2)^2}{4a}] {\rm UnitBox}[\frac{\pi(t_1-t_2)}{2b}]{\rm sin}[\pi(t_1-t_2)\Delta].
\end{equation}
The JTI in Fig.\,\ref{FigS-jsti}d is plotted using Eq.\,\ref{eqS8} with the parameter of $\Delta=0.4237$,
and Fig.\,\ref{FigS-jsti}f is plotted by $\Delta=0.2131$.
Similarly, for two spectral modes with a relative phase of 0 (sum frequency), the JSA is:
\begin{equation}\label{eqS9}
JSA_+(\nu_1, \nu_2)=\exp[-a(\nu_1+\nu_2)^2] \{ {\rm sinc} [b(\nu_1-\nu_2+\Delta)]+ {\rm sinc} [b(\nu_1-\nu_2-\Delta)]\}.
\end{equation}
The JTA is
\begin{equation}\label{eqS10}
JTA_+(t_1, t_2)=\frac{\pi}{b}\sqrt{\frac{\pi}{a}} \exp[-\frac{\pi^2(t_1+t_2)^2}{4a}]  {\rm UnitBox} [\frac{\pi(t_1-t_2)}{2b}]{\rm cos}[\pi(t_1-t_2)\Delta].
\end{equation}
JTI in Fig.\,\ref{FigS-jsti}g is plotted using Eq.\,(\ref{eqS10}).
%
%
%

It is noteworthy that the spectral mode separation in JSI is $\sqrt 2 \Delta$, while the temporal mode separation in JSI is $\frac{1}{{\sqrt 2 \Delta }}$.
Therefore, the mode separation product (MSP) is 1 in the level of intensity. The MSP value is 2 at the level of amplitude. This is the same as in the case of a double-slit experiment.


\begin{thebibliography}{99}
\newcommand{\enquote}[1]{``#1''}

\bibitem{Weiner2009book}
A.~Weiner, \emph{Ultrafast Optics} (Wiley, 2009), 1st ed.

\bibitem{Cundiff2001}
S.~T. Cundiff, J.~Ye, and J.~L. Hall, \enquote{Optical frequency synthesis
  based on mode-locked lasers,} Rev. Sci. Instrum. \textbf{72}, 3749--3771
  (2001).

\bibitem{Jiang2005}
Z.~Jiang, D.~Leaird, and A.~Weiner, \enquote{Line-by-line pulse shaping control
  for optical arbitrary waveform generation,} Opt. Express \textbf{13},
  10431--10439 (2005).

\bibitem{Chan2011}
H.-S. Chan, Z.-M. Hsieh, W.-H. Liang, A.~H. Kung, C.-K. Lee, C.-J. Lai, R.-P.
  Pan, and L.-H. Peng, \enquote{Synthesis and measurement of ultrafast
  waveforms from five discrete optical harmonics,} Science \textbf{331}, 1165
  (2011).

\bibitem{Spencer2018}
D.~T. Spencer, T.~Drake, T.~C. Briles, J.~Stone, L.~C. Sinclair, C.~Fredrick,
  Q.~Li, D.~Westly, B.~R. Ilic, A.~Bluestone, N.~Volet, T.~Komljenovic,
  L.~Chang, S.~H. Lee, D.~Y. Oh, M.-G. Suh, K.~Y. Yang, M.~H.~P. Pfeiffer,
  T.~J. Kippenberg, E.~Norberg, L.~Theogarajan, K.~Vahala, N.~R. Newbury,
  K.~Srinivasan, J.~E. Bowers, S.~A. Diddams, and S.~B. Papp, \enquote{An
  optical-frequency synthesizer using integrated photonics,} Nature
  \textbf{557}, 81--85 (2018).

\bibitem{Giorgetta2010}
F.~R. Giorgetta, I.~Coddington, E.~Baumann, W.~C. Swann, and N.~R. Newbury,
  \enquote{Fast high-resolution spectroscopy of dynamic continuous-wave laser
  sources,} Nature Photon. \textbf{4}, 853 (2010).

\bibitem{Kato2017SRminoshima}
T.~Kato, M.~Uchida, and K.~Minoshima, \enquote{No-scanning 3d measurement
  method using ultrafast dimensional conversion with a chirped optical
  frequency comb,} Sci. Rep. \textbf{7}, 3670 (2017).

\bibitem{Asahara2019APEminoshima}
A.~Asahara and K.~Minoshima, \enquote{Coherent multi-comb pulse control
  demonstrated in polarization-modulated dual-comb spectroscopy technique,}
  Appl. Phys. Express \textbf{12}, 072014 (2019).

\bibitem{Peer2005}
A.~Pe'er, B.~Dayan, A.~A. Friesem, and Y.~Silberberg, \enquote{Temporal shaping
  of entangled photons,} Phys. Rev. Lett. \textbf{94}, 073601 (2005).

\bibitem{Jin2018PRAppl}
R.-B. Jin, T.~Saito, and R.~Shimizu, \enquote{Time-frequency duality of
  biphotons for quantum optical synthesis,} Phys. Rev. Appl. \textbf{10},
  034011 (2018).

\bibitem{MacLean2018}
J.-P.~W. MacLean, J.~M. Donohue, and K.~J. Resch, \enquote{Direct
  characterization of ultrafast energy-time entangled photon pairs,} Phys. Rev.
  Lett. \textbf{120}, 053601 (2018).

\bibitem{Ansari2018}
V.~Ansari, E.~Roccia, M.~Santandrea, M.~Doostdar, C.~Eigner, L.~Padberg,
  I.~Gianani, M.~Sbroscia, J.~M. Donohue, L.~Mancino, M.~Barbieri, and
  C.~Silberhorn, \enquote{Heralded generation of high-purity ultrashort single
  photons in programmable temporal shapes,} Opt. Express \textbf{26},
  2764--2774 (2018).

\bibitem{Ansari2018Optica}
V.~Ansari, J.~M. Donohue, B.~Brecht, and C.~Silberhorn, \enquote{Tailoring
  nonlinear processes for quantum optics with pulsed temporal-mode encodings,}
  Optica \textbf{5}, 534--550 (2018).

\bibitem{Davis2018arXiv}
A.~O. Davis, V.~Thiel, and B.~J. Smith, \enquote{Measuring the quantum state of
  a photon pair entangled in frequency and time,} arXiv:1809.03727  (2018).

\bibitem{Graffitti2019arXiv}
F.~Graffitti, P.~Barrow, A.~Pickston, A.~M. Branczyk, and A.~Fedrizzi,
  \enquote{Direct generation of tailored pulse-mode entanglement,}
  arXiv:1910.00598  (2019).

\bibitem{Branning1999}
D.~Branning, W.~P. Grice, R.~Erdmann, and I.~A. Walmsley, \enquote{Engineering
  the indistinguishability and entanglement of two photons,} Phys. Rev. Lett.
  \textbf{83}, 955--958 (1999).

\bibitem{Jin2018OE}
R.-B. Jin, R.~Shiina, and R.~Shimizu, \enquote{Quantum manipulation of biphoton
  spectral distributions in a {2D} frequency space toward arbitrary shaping of
  a biphoton wave packet,} Opt. Express \textbf{26}, 21153--21158 (2018).

\bibitem{Grice2001}
W.~P. Grice, A.~B. U'Ren, and I.~A. Walmsley, \enquote{Eliminating frequency
  and space-time correlations in multiphoton states,} Phys. Rev. A \textbf{64},
  063815 (2001).

\bibitem{Eckstein2011}
A.~Eckstein, A.~Christ, P.~J. Mosley, and C.~Silberhorn, \enquote{Highly
  efficient single-pass source of pulsed single-mode twin beams of light,}
  Phys. Rev. Lett. \textbf{106}, 013603 (2011).

\bibitem{Cui2019PRAppl}
C.~Cui, R.~Arian, S.~Guha, N.~Peyghambarian, Q.~Zhuang, and Z.~Zhang,
  \enquote{Wave-function engineering for spectrally uncorrelated biphotons in
  the telecommunication band based on a machine-learning framework,} Phys. Rev.
  Appl. \textbf{12}, 034059 (2019).

\bibitem{Shimizu2009}
R.~Shimizu and K.~Edamatsu, \enquote{High-flux and broadband biphoton sources
  with controlled frequency entanglement,} Opt. Express \textbf{17},
  16385--16393 (2009).

\bibitem{Gerrits2015}
T.~Gerrits, F.~Marsili, V.~B. Verma, L.~K. Shalm, M.~Shaw, R.~P. Mirin, and
  S.~W. Nam, \enquote{Spectral correlation measurements at the {Hong-Ou-Mandel}
  interference dip,} Phys. Rev. A \textbf{91}, 013830 (2015).

\bibitem{Jin2016QST}
R.-B. Jin, R.~Shimizu, M.~Fujiwara, M.~Takeoka, R.~Wakabayashi, T.~Yamashita,
  S.~Miki, H.~Terai, T.~Gerrits, and M.~Sasaki, \enquote{Simple method of
  generating and distributing frequency-entangled qudits,} Quantum Sci.
  Technol. \textbf{1}, 015004 (2016).

\bibitem{Kuzucu2008PRL}
O.~Kuzucu, F.~N.~C. Wong, S.~Kurimura, and S.~Tovstonog, \enquote{Joint
  temporal density measurements for two-photon state characterization,} Phys.
  Rev. Lett. \textbf{101}, 153602 (2008).

\bibitem{Hong1987}
C.~K. Hong, Z.~Y. Ou, and L.~Mandel, \enquote{Measurement of subpicosecond time
  intervals between two photons by interference,} Phys. Rev. Lett. \textbf{59},
  2044--2046 (1987).

\bibitem{Lu2018OL}
H.-H. Lu, O.~D. Odele, D.~E. Leaird, and A.~M. Weiner, \enquote{Arbitrary
  shaping of biphoton correlations using near-field frequency-to-time mapping,}
  Opt. Lett. \textbf{43}, 743--746 (2018).

\bibitem{Schmitt2007}
M.~Schmitt, B.~Dietzek, G.~Hermann, and J.~Popp, \enquote{Femtosecond
  time-resolved spectroscopy on biological photoreceptor chromophores,} Laser
  Photonics Rev. \textbf{1}, 57--78 (2007).

\bibitem{Maiuri2020}
M.~Maiuri, M.~Garavelli, and G.~Cerullo, \enquote{Ultrafast spectroscopy: State
  of the art and open challenges,} J. Am. Chem. Soc. \textbf{142}, 3--15
  (2020).

\bibitem{Mukamel2020JPB}
S.~Mukamel, M.~Freyberger, W.~P. Schleich, M.~Bellini, A.~Zavatta, G.~Leuchs,
  C.~Silberhorn, R.~W. Boyd, L.~S. Soto, A.~Stefanov, M.~Barbieri, A.~Paterova,
  L.~A. Krivitskiy, S.~Shwartz, K.~Tamasaku, K.~Dorfman, F.~Schlawin,
  V.~Sandoghdar, M.~G. Raymer, A.~H. Marcus, O.~Varnavski, T.~Goodson~III,
  Z.-Y. Zhou, B.-S. Shi, S.~Asban, M.~O. Scully, G.~S. Agarwal, T.~Peng, A.~V.
  Sokolov, Z.~Zhang, I.~A. Vartaniants, E.~del Valle, and F.~P. Laussy,
  \enquote{Roadmap on quantum light spectroscopy,} J. Phys. B: At. Mol. Opt.
  Phys. (in press)  (2020).

\bibitem{Schlawin2013}
F.~Schlawin, K.~E. Dorfman, B.~P. Fingerhut, and S.~Mukamel,
  \enquote{Suppression of population transport and control of exciton
  distributions by entangled photons,} Nature Commun. \textbf{4}, 1782 (2013).

\bibitem{Fujihashi2019arXiv}
Y.~Fujihashi, R.~Shimizu, and A.~Ishizaki, \enquote{Generation of
  pseudo-sunlight via quantum entangled photons and the interaction with
  molecules,} arXiv:1904.11669  (2019).

\bibitem{Xu2019arXiv}
F.~Xu, X.~Ma, Q.~Zhang, H.-K. Lo, and J.-W. Pan, \enquote{Secure quantum key
  distribution with realistic devices,} arXiv:1903.09051  (2019).

\bibitem{Miki2013}
S.~Miki, T.~Yamashita, H.~Terai, and Z.~Wang, \enquote{{High performance
  fiber-coupled NbTiN superconducting nanowire single photon detectors with
  Gifford-McMahon cryocooler},} Opt. Express \textbf{21}, 10208--10214 (2013).

\end{thebibliography}
\end{document}